# Generalized relativistic velocity addition with spacetime algebra

C R Paiva and M A Ribeiro

Instituto de Telecomunicações, Instituto Superior Técnico, Av. Rovisco Pais, 1, 1049-001 Lisboa, Portugal

E-mail: c.paiva@ieee.org

**Abstract**

Using spacetime algebra – the geometric algebra of spacetime – the general problem of relativistic addition of velocities is addressed. The successive application of non-collinear Lorentz boosts is then studied in Minkowski spacetime. Even spatial vectors, such as the relative velocity of two reference frames, are analyzed in their proper setting – as vectors belonging to a four-dimensional spacetime manifold and not as vectors in ordinary three-dimensional (Euclidean) space. This is a clear and physical result that illustrates the fact that Lorentz boosts do not form a group. The entire derivation is carried through without using Einstein's second postulate on the speed of light, thereby stressing that the framework of special relativity does not depend on electromagnetism.

**1. Introduction**

The most dangerous of pre-relativistic prejudices is the belief that there is an absolute meaning to the simultaneity of two events that occur at two different places, i.e., independently of the reference frame in which those events are described. That is why special relativity should be taught in high school: this kind of pre-relativistic prejudices can (and should) be demystified as early as possible, using the most elementary mathematical approach available. Perhaps the excellent recent book from David Mermin [1] can help along this quest. However, at the (senior) undergraduate level, special relativity should be developed using the most appropriate mathematical tools. A typical example where a more sophisticated mathematical approach is needed is the problem of relativistic addition of velocities in its more general form. In fact, if two successive boosts are non-collinear then the combined transformation is the composite of a rotation and a single boost. In other words, boosts do not form a group. This was first discovered by Thomas [2] and extensively analyzed by Wigner [3]. In this paper we present a fresh look at this problem using spacetime algebra [4].

It is now commonly recognized that it is possible to derive the Lorentz transformation without Einstein's second postulate which states that the speed of light $c$ (in a vacuum) is the same for all inertial (non-accelerating) observers [5-8]. This new type of approach has, from the epistemological point of view, deep consequences: it shows that special relativity is independent from electromagnetism, thereby establishing its universal applicability – as far as gravitation is not concerned.

On the other hand, it is the authors' opinion that geometric algebra [9] provides the best mathematical framework to deal with physical problems in Minkowski spacetime. The traditional approach based on four-vectors [10] treats spatial vectors as entities in three-dimensional Euclidean space, not as vectors belonging to spacetime associated with a given observer. However, in [9] only the relativistic addition of velocities for the particular case of collinear boosts is studied. The only treatment of the general problem of non-collinear boosts with geometric algebra can be found in [11]. Nevertheless, in [11] the approach to Minkowski spacetime is different from the one adopted in [4] and [9]: instead of using spacetime algebra as a framework to express an invariant formulation of physics – as opposed to relativistic approaches, where the major concern is to interpret results according to different observers – every event in spacetime is encoded through a paravector (the sum of a scalar and a three-dimensional vector from ordinary Euclidean space). Accordingly, the relativistic approach presented in [11] is physically equivalent to traditional four-vectors as it cannot present the physical variables in their natural setting – Minkowski spacetime modelled by spacetime algebra as in [4]. The spatial component of traditional four-vectors, which corresponds indeed to a spatial vector in ordinary space, should not be dissociated from the proper time of a specific observer. That is why in this paper we present the generalized relativistic velocity addition using the spacetime algebra of [4] and [9]. However, in order to get a more sound approach from the epistemological perspective, we also adopt the heterodox viewpoint that disregards any consideration from electromagnetic theory (i.e., without Einstein's second postulate on the speed of light). We hope that this paper can help to bridge the gap between relativistic physics and proper spacetime physics by shedding light onto the physics behind the mathematical formalism of spacetime algebra.

## 2. Proper and relative velocities

Probably the most important mathematical concept in special relativity is the correspondence between an *event* in spacetime and its representation as a vector, say vector $a$, in the four-dimensional (flat) manifold that we shall call Minkowski spacetime $M$. Let $\{e_0, e_1, e_2, e_3\}$ be an orthogonal basis $S$ for $M$ and $e_0$ the unit vector for the time axis, with $e_0^2 = 1$, so that

$$a = (\kappa t)e_0 + \mathbf{r}, \quad \mathbf{r} = x e_1 + y e_2 + z e_3. \tag{1}$$

Here $\kappa$ is just an appropriate constant (with dimension of velocity) that transforms units of time into units of length; no other meaning is ascribed *a priori* to it. One should note that $\mathbf{r} \in M$ is a spatial vector (henceforth all spatial vectors will be written in boldface type) that is orthogonal to $e_0$. However, at this moment, we cannot decide whether $e_1^2 = e_2^2 = e_3^2 = 1$ or $e_1^2 = e_2^2 = e_3^2 = -1$. The same event $a$ from (1) is represented, in another frame $\bar{S}$, by a new basis $\{f_0, f_1, f_2, f_3\}$ with

$$a = (\kappa \bar{t})f_0 + \bar{\mathbf{r}}, \quad \bar{\mathbf{r}} = \bar{x} f_1 + \bar{y} f_2 + \bar{z} f_3. \tag{2}$$

Obviously, $f_0^2 = 1$ as $e_0^2 = 1$. Moreover, if $e_1^2 = e_2^2 = e_3^2 = 1$ then $f_1^2 = f_2^2 = f_3^2 = 1$; however, if $e_1^2 = e_2^2 = e_3^2 = -1$ then $f_1^2 = f_2^2 = f_3^2 = -1$.

A rest point $P$ in reference frame $S$ will be represented by the world line $a_P(t) = (\kappa t)e_0 + \mathbf{r}_P$ where $\mathbf{r}_P$ is a constant vector (it does not depend on proper time $t$). In frame $\bar{S}$ the same rest point will be represented by the world line $a_P(\bar{t}) = (\kappa \bar{t})f_0 + \bar{\mathbf{r}}_P(\bar{t})$. Hence, denoting by $s = d a_P / dt = (d a_P / d\bar{t})(d\bar{t}/dt)$ the proper velocity of rest point $P$, we will have

$$s = \kappa \mathrm{e}_0 = \gamma(\kappa \mathrm{f}_0 + \mathbf{w}), \quad \mathbf{w} = \frac{d\overline{\mathbf{r}}_P}{d\overline{t}} \qquad (3)$$

where $\overline{t}$ is the proper time of $\overline{S}$ and $\gamma = d\overline{t}/dt$. Similarly, for a rest point $Q$ in $\overline{S}$ the corresponding world line will be $a_Q(\overline{t}) = (\kappa \overline{t})\mathrm{f}_0 + \overline{\mathbf{r}}_Q$ for an observer in $\overline{S}$, whereas $a_Q(t) = (\kappa t)\mathrm{e}_0 + \mathbf{r}_Q(t)$ for an observer in $S$. Therefore, the proper velocity of rest point $Q$ will be $u = da_Q/d\overline{t} = (da_P/dt)(dt/d\overline{t})$ and hence

$$u = \kappa \mathrm{f}_0 = \overline{\gamma}(\kappa \mathrm{e}_0 + \mathbf{v}), \quad \mathbf{v} = \frac{d\mathbf{r}_Q}{dt} \qquad (4)$$

where $\overline{\gamma} = dt/d\overline{t}$. However, according to the principle of relativity, one should have $\gamma = \overline{\gamma}$ in (3) and (4). But then

$$\begin{cases} s = \kappa \mathrm{e}_0 = \gamma(u + \mathbf{w}) \\ u = \kappa \mathrm{f}_0 = \gamma(s + \mathbf{v}) \end{cases}. \qquad (5)$$

Apart from proper velocities $s$ and $u$ we also have relative velocities $\mathbf{v}$ and $\mathbf{w}$: whereas $\mathbf{v}$ is the velocity that $\overline{S}$ is moving away from the perspective of an observer in $S$, $\mathbf{w}$ is the velocity that $S$ is moving away from the perspective of an observer in $\overline{S}$. However, the magnitudes of both $\mathbf{v}$ and $\mathbf{w}$ should be the same. By convention we make

$$\begin{cases} \mathbf{v} = \beta \kappa \hat{\mathbf{v}} \\ \mathbf{w} = -\beta \kappa \hat{\mathbf{w}} \end{cases} \qquad (6)$$

where $\hat{\mathbf{v}}$ and $\hat{\mathbf{w}}$ are unit vectors with either $\hat{\mathbf{v}}^2 = \hat{\mathbf{w}}^2 = 1$ or $\hat{\mathbf{v}}^2 = \hat{\mathbf{w}}^2 = -1$ (to be determined later). From (5) and (6), we get

$$\begin{cases} \mathrm{e}_0 = \gamma(\mathrm{f}_0 - \beta \hat{\mathbf{w}}) \\ \mathrm{f}_0 = \gamma(\mathrm{e}_0 + \beta \hat{\mathbf{v}}) \end{cases} \qquad (7)$$

and hence

$$\begin{cases} \hat{\mathbf{v}} = \gamma\left(\hat{\mathbf{w}} - \frac{\gamma^2 - 1}{\gamma^2 \beta}\mathrm{f}_0\right) \\ \hat{\mathbf{w}} = \gamma\left(\hat{\mathbf{v}} + \frac{\gamma^2 - 1}{\gamma^2 \beta}\mathrm{e}_0\right) \end{cases}. \qquad (8)$$

## 3. Geometric product of vectors

In this paper we will use the geometric *product* of vectors: if $a, b \in M$ then $(a,b) \mapsto ab$. This geometric product should reduce to $ab = a \cdot b$ whenever vectors $a$ and $b$ are parallel. Furthermore, it should reduce to $ab = a \wedge b$ whenever vectors $a$ and $b$ are orthogonal. As usual $a \cdot b$ is the *inner* (or dot) product which is a real number; $a \wedge b$ is the *outer* (or exterior) product which is, in this case (outer product of two vectors), a *bivector* which encodes an oriented plane in spacetime $M$. The inner product is symmetric $(a \cdot b = b \cdot a)$, whereas the outer product is antisymmetric $(a \wedge b = -b \wedge a)$. Then, in general, we define the geometric product as

$$ab = a \cdot b + a \wedge b. \qquad (9)$$

We see, from (9), that the result of the geometric product is not – in general – either a scalar or a bivector: it is what we shall call a *multivector*. Indeed, the right-hand side of (9) should be considered as a graded sum: the sum of two objects of different *grade* (as the sum of a real number with an imaginary number). Multivectors can be broken up into terms of different grade: the scalar part is assigned grade 0, the vector grade 1 and bivectors grade 2. We denote the projection onto terms of a chosen grade $k$ by $\langle \ \rangle_k$, so that in (9) we have $a \cdot b = \langle ab \rangle_0$ and $a \wedge b = \langle ab \rangle_2$. The reverse of multivector (9) is defined as

$$(ab)^\dagger = ba = a \cdot b - a \wedge b. \tag{10}$$

The geometric product between vectors is associative, (left and right) distributive over addition and obeys to the so-called contraction: $a^2 = aa \in \mathbb{R}$. It is this last property that distinguishes the graded geometric algebra from a general associative algebra (e.g., matrix algebra). We do not force the contraction to be positive – just a real number. That is why we establish, by convention, that

$$e_0^2 = f_0^2 = 1. \tag{11}$$

However, only later – based upon physical arguments – we will establish the sign of $\hat{\mathbf{v}}^2 = \hat{\mathbf{w}}^2$ as $+1$ or as $-1$. According to (9) and (10), we get

$$\begin{cases} a \cdot b = \dfrac{1}{2}\left[ab + (ab)^\dagger\right] = \dfrac{1}{2}(ab + ba) \\ a \wedge b = \dfrac{1}{2}\left[ab - (ab)^\dagger\right] = \dfrac{1}{2}(ab - ba) \end{cases}. \tag{12}$$

One should note, from $(a+b)^2 = (ab - a \cdot b)(a \cdot b - ba) = (a \cdot b)(ab + ba) - a^2 b^2 - (a \cdot b)^2$, that

$$(a \wedge b)^2 = (a \cdot b)^2 - a^2 b^2 \tag{13}$$

and hence $(a \wedge b)^2 \in \mathbb{R}$.

Now, using the geometric product of vectors is spacetime, we get from (7)

$$f_0 e_0 = \gamma(1 + \beta \hat{B}), \quad \hat{B} = \hat{\mathbf{v}} e_0. \tag{14}$$

Unit bivector $\hat{B} = \hat{\mathbf{v}} e_0 = \hat{\mathbf{v}} \wedge e_0$ characterizes the relative velocity between frames $S$ and $\bar{S}$ as $\hat{\mathbf{v}} = \hat{B} e_0$. Then, according to the principle of relativity, one should have $\hat{\mathbf{w}} = \hat{B} f_0$ and hence we define the relative velocity though bivector $B$ given by

$$B = \beta \hat{B}, \quad \hat{B} = \hat{\mathbf{v}} e_0 = \hat{\mathbf{w}} f_0 \tag{15}$$

where

$$\hat{B}^2 = (\hat{\mathbf{v}} e_0)(\hat{\mathbf{v}} e_0) = -\hat{\mathbf{v}} e_0 e_0 \hat{\mathbf{v}} = -\hat{\mathbf{v}}^2 = -\hat{\mathbf{w}}^2. \tag{16}$$

Whence, it also follows that either $\hat{B}^2 = 1$ or $\hat{B}^2 = -1$. Defining a multivector $L$, such that

$$L^2 = f_0 e_0 = f_0 \cdot e_0 + f_0 \wedge e_0, \tag{17}$$

we obtain, from (14) and (17),

$$\begin{cases} f_0 \cdot e_0 = \gamma \\ f_0 \wedge e_0 = \gamma \beta \hat{B} \end{cases} \quad \therefore \quad L^2 = \gamma(1 + \beta \hat{B}). \tag{18}$$

Generalized relativistic velocity addition with spacetime algebra

Using (13) and (18), then

$$(f_0 \wedge e_0)^2 = \gamma^2 \beta^2 \hat{B}^2 = \gamma^2 - 1 \quad \therefore \quad \gamma = \frac{1}{\sqrt{1-\beta^2 \hat{B}^2}}. \tag{19}$$

Therefore, from (8), we also get

$$\frac{\gamma^2 - 1}{\gamma^2 \beta} = \beta \hat{B}^2 \quad \therefore \quad \begin{cases} \hat{\mathbf{v}} = \gamma(\hat{\mathbf{w}} - \beta \hat{B}^2 f_0) \\ \hat{\mathbf{w}} = \gamma(\hat{\mathbf{v}} + \beta \hat{B}^2 e_0) \end{cases}. \tag{20}$$

One should note that, using (7) and (20), we can in fact obtain, in accordance with (15), $\hat{B} = \hat{\mathbf{v}} e_0 = \gamma^2(1 - \beta^2 \hat{B}^2)(\hat{\mathbf{w}} f_0) = \hat{\mathbf{w}} f_0$. The reverse of bivector $\hat{B}$ is $\hat{B}^\dagger = e_0 \hat{\mathbf{v}} = -\hat{B}$ so that, according to (18) and (19),

$$(L^2)^\dagger = \gamma(1 - \beta \hat{B}), \quad L^2 (L^2)^\dagger = 1. \tag{21}$$

If multivector $L$ also obeys to $LL^\dagger = 1$ (we say that a multivector that satisfies this relation is a *rotor*) then $L^2 + 1 = (L + L^\dagger)L$. On the other hand, as $L + L^\dagger = 2\langle L \rangle_0$, we obtain from (18)

$$L = \frac{1 + L^2}{2\langle L \rangle_0} = \frac{1 + \gamma + \gamma \beta \hat{B}}{\sqrt{2(1+\gamma)}} \tag{22}$$

and hence

$$L e_0 = e_0 L^\dagger, \quad L \hat{\mathbf{v}} = \hat{\mathbf{v}} L^\dagger. \tag{23}$$

One can easily see that (7) and (20) can be rewritten in the more symmetrical form

$$f_0 = L e_0 L^\dagger, \quad \hat{\mathbf{w}} = L \hat{\mathbf{v}} L^\dagger. \tag{24}$$

We call the transformation $L: S \to \bar{S}; e_0 \mapsto f_0, \hat{\mathbf{v}} \mapsto \hat{\mathbf{w}}$ characterized by rotor $L$ (we use the same letter for the transformation and for the rotor) a *boost*. It is an active Lorentz transformation. In figure 1 the Lorentz transformation $L$ between frames $S$ and $\bar{S}$ is schematically shown.

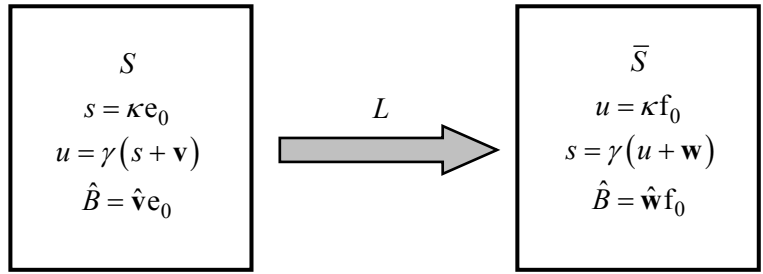

**Figure 1.** The Lorentz transformation between frames $S$ and $\bar{S}$. An observer (in rest) in $S$ has proper velocity $s$ and sees $\bar{S}$ receding $(\beta > 0)$ or approaching $(\beta < 0)$ with relative velocity $\mathbf{v} = \beta \kappa \hat{\mathbf{v}}$, whereas an observer in $\bar{S}$ has proper velocity $u = L s L^\dagger$ and sees $S$ moving away with relative velocity $\mathbf{w} = -\beta \kappa \hat{\mathbf{w}}$. Bivector $B = \beta \hat{B}$ is null when there is no relative motion $(\beta = 0)$.

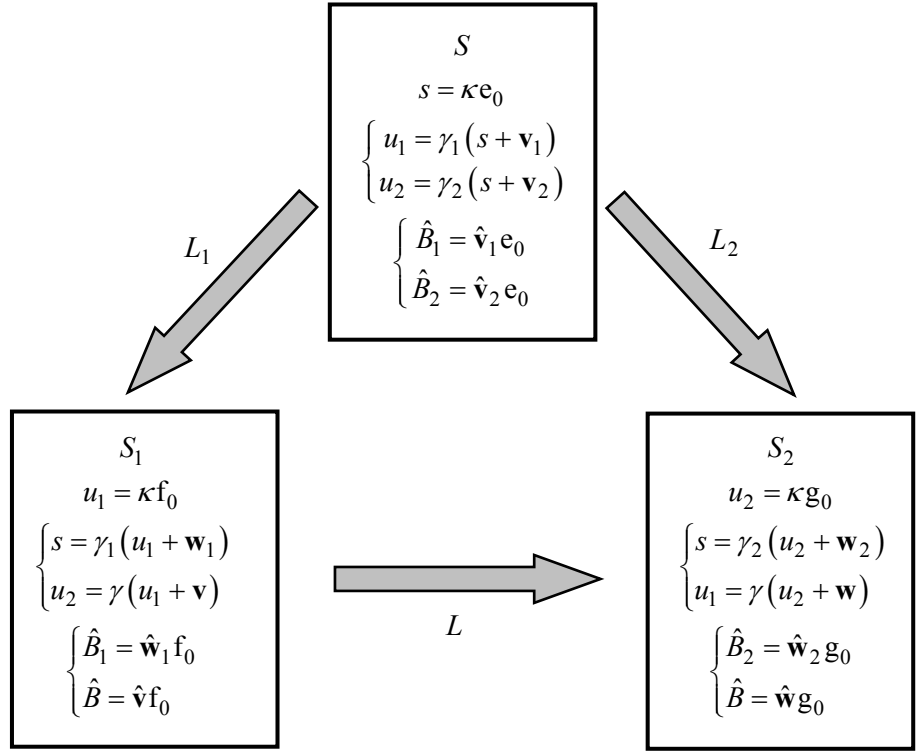

**Figure 2.** The composition of two non-collinear Lorentz boosts $L_1$ and $L_2$. An observer (in rest) in frame $S$, with proper velocity $s$, sees observers $S_1$ and $S_2$ moving away with relative velocities $\mathbf{v}_1$ and $\mathbf{v}_2$ that are not parallel in general (bivectors $\hat{B}_1$ and $\hat{B}_2$ do not coincide). Observer $S_1$ (with proper velocity $u_1$) sees observer $S_2$ (with proper velocity $u_2$) moving away with relative velocity $\mathbf{v}$. Rotor $L$ is associated with bivector $\hat{B}$ which, in the general case, does not coincide either with $\hat{B}_1$ or with $\hat{B}_2$.

## 4. Addition of velocities in Minkowski spacetime

The main goal of this paper is to analyze the composition of non-collinear Lorentz boosts. That is the objective of this section as shown is figure 2.

There are three different observers: (i) an observer (in rest) in reference frame $S$ with proper velocity $s = \kappa e_0$; (ii) an observer $S_1$ with proper velocity $u_1 = \kappa f_0$; (iii) an observer $S_2$ with proper velocity $u_2 = \kappa g_0$. Observer $S$ sees observers $S_1$ and $S_2$ moving away with relative velocities $\mathbf{v}_1 = \beta_1 \kappa \hat{\mathbf{v}}_1$ and $\mathbf{v}_2 = \beta_2 \kappa \hat{\mathbf{v}}_2$, respectively; observer $S_1$, on the other hand, sees observer $S_2$ moving away with relative velocity $\mathbf{v} = \beta \kappa \hat{\mathbf{v}}$. The physical problem can be put as follows: given $\mathbf{v}_1$ and $\mathbf{v}_2$ one wishes to determine $\mathbf{v}$. However, one should note that we are working in Minkowski spacetime: vectors $\mathbf{v}_1$, $\mathbf{v}_2$ and $\mathbf{v}$ belong to different observers and hence one should take into account the fact that these different observers have different proper times.

The solution for our problem is, according to figure 2, obvious: as $u_2 = \gamma(u_1 + \mathbf{v})$, one has

$$\mathbf{v} = \frac{u_2}{\gamma} - u_1. \tag{25}$$

Then it is only necessary to know $\gamma$ in terms of $\gamma_1$ and $\gamma_2$. From

$$\begin{cases} u_1 = \kappa f_0 = \gamma_1(s + \mathbf{v}_1) \\ u_2 = \kappa g_0 = \gamma_2(s + \mathbf{v}_2) \end{cases} \quad (26)$$

we get, according to (18) and from $\mathbf{v}_1 \cdot e_0 = \mathbf{v}_2 \cdot e_0 = 0$,

$$\begin{cases} u_1 \cdot u_2 = \kappa^2 (f_0 \cdot g_0) = \kappa^2 \gamma \\ u_1 \cdot u_2 = \gamma_1 \gamma_2 (\kappa^2 + \mathbf{v}_1 \cdot \mathbf{v}_2) \end{cases}.$$

Therefore,

$$\gamma = \gamma_1 \gamma_2 \left[1 + \beta_1 \beta_2 (\hat{\mathbf{v}}_1 \cdot \hat{\mathbf{v}}_2)\right]. \quad (27)$$

In general one should write

$$\hat{\mathbf{v}}_1 \cdot \hat{\mathbf{v}}_2 = -\hat{B}^2 \cos\theta. \quad (28)$$

Only when $\hat{B}^2 = -1$, do we recover a Euclidean metric for spacetime (if that is the case). But then, using (25)-(28) and

$$\hat{\mathbf{w}}_1 = \gamma_1 \left(\hat{\mathbf{v}}_1 + \beta_1 \hat{B}^2 e_0\right),$$

we finally get

$$\mathbf{v} = \kappa \frac{\beta_2 (\hat{\mathbf{v}}_2 - \hat{\mathbf{v}}_1 \cos\theta) + \gamma_1 (\beta_2 \cos\theta - \beta_1) \hat{\mathbf{w}}_1}{\gamma_1 \left(1 - \beta_1 \beta_2 \hat{B}^2 \cos\theta\right)}. \quad (29)$$

There is only one problem that remains unsolved: we must know whether $\hat{B}^2 = -1$ or $\hat{B}^2 = 1$. Let us consider the particular case corresponding to $\hat{\mathbf{v}}_1 = \hat{\mathbf{v}}_2$ ($\theta = 0$). From (29) we get $\mathbf{v} = \beta \kappa \hat{\mathbf{w}}_1$ (i.e., $\hat{B}_1 = \hat{B}_2 = \hat{B}$ in figure 2), with

$$\beta = \frac{\beta_2 - \beta_1}{1 - \beta_1 \beta_2 \hat{B}^2}. \quad (30)$$

If $\hat{B}^2 = -1$ we obtain a physically strange result: for $\beta_1 = 3$ and $\beta_2 = 1$ we get $\beta = -1/2$, i.e., the (relativistic) addition of two velocities in the same direction results in a velocity in the opposite direction. This suggests that, to avoid these strange results, we should always adopt $\hat{B}^2 = 1$. This means that, along with (11), one should always have

$$e_1^2 = e_2^2 = e_3^2 = -1. \quad (31)$$

That is why the metric of Minkowski spacetime is not Euclidean (positive definite) – it is a Lorentz metric (negative indefinite). One should note that, instead of (11), we could have imposed $e_0^2 = f_0^2 = -1$. In that case, however, we would get $e_1^2 = e_2^2 = e_3^2 = 1$ (i.e., also a Lorentz metric although positive indefinite).

## 5. Concluding remarks

Using only the principle of relativity, the generalized relativistic addition of velocities (corresponding to non-collinear Lorentz boosts in the general case) was found:

Generalized relativistic velocity addition with spacetime algebra

$$\mathbf{v} = \kappa \frac{\beta_2(\hat{\mathbf{v}}_2 - \hat{\mathbf{v}}_1 \cos\theta) + \gamma_1(\beta_2 \cos\theta - \beta_1)\hat{\mathbf{w}}_1}{\gamma_1(1 - \beta_1\beta_2 \cos\theta)}. \tag{32}$$

The fact that in general $\hat{B}_1 \neq \hat{B}_2$ (figure 2) means that the composition of two boosts is not a boost; boosts, therefore, do not form a group. Spatial rotations are needed to transform $\hat{B}_1$ into $\hat{B}_2$ (in $S$), $\hat{B}_1$ into $\hat{B}$ (in $S_1$) or $\hat{B}_2$ into $\hat{B}$ (in $S_3$).

Moreover, we have also found that Minkowski spacetime should have a signature corresponding to (11) and (31). This means that, according to (1), one has

$$\begin{cases} a\mathrm{e}_0 = \kappa t + R \\ \mathrm{e}_0 a = \kappa t - R \end{cases} \tag{33}$$

where bivector $R = \mathbf{r}\mathrm{e}_0 = \mathbf{r} \wedge \mathrm{e}_0$ was introduced. Then, we obtain the invariant interval of special relativity

$$a^2 = (a\mathrm{e}_0)(\mathrm{e}_0 a) = \kappa^2 t^2 - R^2 = \kappa^2 t^2 + \mathbf{r}^2 = \kappa^2 t^2 - r^2 \tag{34}$$

where $R = r\hat{R}$ and $\mathbf{r} = r\hat{\mathbf{r}}$, with $\hat{R}^2 = -\hat{\mathbf{r}}^2 = 1$. Likewise, the passive Lorentz transformations can be readily derived from (1) and (2) by taking into account that $\mathbf{r} = \mathbf{r}_\parallel + \mathbf{r}_\perp$ and $\mathbf{r}_\parallel = -(\mathbf{r} \cdot \hat{\mathbf{v}})\hat{\mathbf{v}}$. The special case of collinear boosts $(\theta = 0)$ reduces to

$$\mathbf{v} = \frac{v_2 - v_1}{1 - v_1 v_2 / \kappa^2} \hat{\mathbf{w}}_1 \tag{35}$$

according to (32). From (35) a natural limit $\kappa > 0$ for all particle velocities arises: if $v = \beta\kappa$, one always has $|\beta| \leq 1$ or $|v| \leq \kappa$. The fact that $\kappa = c$, where $c$ is the speed of light in a vacuum, is strictly outside the specific scope of special relativity: it belongs to electromagnetic theory and, ultimately, to experience. The Galilean limit, which corresponds to $\kappa = \infty$, is ruled out by experimental evidence – using, not necessarily, electromagnetic signals. One should note that, in the Galilean limit, one gets from (6) the well-known result $\mathbf{w} = -\mathbf{v}$ as $\beta\kappa = v$ (even if, in this case, $\beta = 0$ and $\kappa = \infty$).

The approach herein presented places the problem of relativistic addition of velocities in the natural setting: the four-dimensional Minkowski spacetime. The framework of geometric algebra, based on the geometric product of vectors, provides the efficient mathematical tools that enable a simple, clear and coordinate-free treatment of vectors in spacetime. Furthermore, with this approach, Einstein's second postulate on the speed of light readily becomes superfluous – something that is, from the epistemological point of view, rather important as it shows that special relativity does not depend on electromagnetism.